\documentclass[twocolumn,showpacs,preprintnumbers,showkeys,amsmath,amssymb]{revtex4}
\usepackage{graphicx}% Include figure files
\usepackage{dcolumn}% Align table columns on decimal point
\usepackage{bm}
\usepackage{natbib}% bold math
\usepackage[T1]{fontenc}
\begin{document}

%\preprint{APS/123-QED}

\title{Magnetic behavior of a non-extensive S-spin system: possible connections to manganites}

\author{M.S. Reis}
\altaffiliation[Also at ]{Centro Brasileiro de Pesquisas Físicas,
Rua Dr. Xavier Sigaud 150 Urca, 22290-180 Rio de Janeiro-RJ,
Brasil} \email{marior@cbpf.br}
\author{J.P. Araújo, V.S. Amaral}
\affiliation{Departamento de Física, Universidade de Aveiro,
3810-192 Aveiro, Portugal}

\author{E.K. Lenzi and I.S. Oliveira}
\affiliation{Centro Brasileiro de Pesquisas Físicas, Rua Dr.
Xavier Sigaud 150 Urca, 22290-180 Rio de Janeiro-RJ, Brasil
}%

\date{\today}

\begin{abstract}
We analyzed the magnetic behavior of a S-spin system within
framework of the Tsallis nonextensive statistics, employing the
normalized approach. Unusual properties on magnetization, entropy
and susceptibility emerge, as a consequence of nonextensivity.  We
further show that the nonextensive approach can be relevant to the
field of manganites, materials which exhibit long-range
interactions and fractality, two basic ingredients for
nonextensivity.  Our results are in qualitative agreement to
experimental data in La$_{0.67}$Ca$_{0.33}$MnO$_3$ and
Pr$_{0.5}$Ca$_{0.5}$Mn$_{0.95}$Ga$_{0.05}$O$_3$ manganites.
\end{abstract}

%\pacs{PACS...}
 \maketitle

\section{Introduction}
In 1988, Tsallis\cite{JSP_52_1988_479} proposed a $q$-dependent
entropy functional which generalized the standard
Maxwell-Boltzmann definition, to include non-extensive systems. In
this formalism, "$q$" is called the \emph{entropic index}, and
measures the degree of nonextensivity of the system.  A few years
latter, Curado and Tsallis\cite{JPAMG_24_1991_L69} revised the
formalism introducing the \emph{unormalized} constraint to the
internal energy and showed that from this entropy functional a
nonextensive thermodynamics could be derived. This generalization
included the classical thermodynamics (Maxwell-Boltzmann), which
could  be recovered when $q$ is set to unity.

After Curado and Tsallis work, this formalism has been
successfully applied to various physical systems, where
Maxwell-Boltzmann framework fails. These include: self-gravitating
systems\cite{PLA_174_1993_384,proceeding1},
turbulence\cite{PRE_61_2000_3237,{JPA_33_2000_L235},{PA_277_2000_115},{PRE_63_2001_035303}},
anomalous
diffusion\cite{{PA_222_1995_347},{PRE_54_1996_R2197},{PRE_62_2000_2213},{PRE_63_2001_30101R}},
velocities of galaxies\cite{ALC_35/6_1998_449}, solar
neutrinos\cite{NPB_87_2000_209}, etc.

In spite of these successes, some drawbacks were identified in the
early formalism, namely: (a) the density operator was not
invariant under a uniform translation of energy spectrum; (b) the
$q$-expected value of the identity operator was not the unity; (c)
energy was not conserved\cite{PA_261_1998_534}. In a more recent
work, Tsallis et al.\cite{PA_261_1998_534} circumvent these
difficulties introducing the \emph{normalized} constraint to the
internal energy of the system.

Concerning applications to magnetic systems, the unormalized
formalism was used by Portesi et al.\cite{PRE_52_1995_R3317} and
Nobre et al.\cite{PMB_73_1996_754}, to describe the paramagnetic
behavior of a system with N spins $1/2$. The authors found a
non-measurable magnetic susceptibility, since it was exponentially
dependent on the number N of particles. Martínez et
al.\cite{PA_282_2000_193} analyzed the same S=1/2 system in the
framework of the normalized formalism\cite{PA_261_1998_534},
finding a similar result to that reported by the Portesi et al.
and Nobre et al.

In the present work, we analyzed the paramagnetic behavior of N
spins S, within the normalized formalism and showed that the
system effective temperature T does not relate with the inverse of
the Lagrange multiplier $\beta$, as it is normally assumed, but
with the inverse of a re-scaled parameter $\beta^*$ ($T=1/k
\beta^*$), already introduced by Tsallis\cite{PA_261_1998_534}. In
this approach, the unreal result found by
Portesi\cite{PRE_52_1995_R3317}, Nobre\cite{PMB_73_1996_754} and
Martínez\cite{PA_282_2000_193} was no longer found and the
susceptibility became proportional to number of particles, as it
is experimentally expected. In these early works, no attempt was
made to correlate theoretical results to experimental magnetic
systems of any kind.

We also suggest that the manganites are physical systems where the
present concepts can be tested. Experimental data of Amaral et
al.\cite{JMMM_226_2001_837} and Hébert et
al.\cite{JSSC_165_2002_6}, are in qualitative agreement with the
results here reported. An analysis of non-extensivity in
manganites in the ferromagnetic phase, also taking $T=1/k \beta^*$
as an effective temperature, was already published
elsewhere\cite{EL_58_2002_42}.

\section{The Model}

We consider the Hamiltonian of a single spin S in a static and
homogeneous magnetic field:

\begin{equation}\label{hamiltoniana}
  \hat{\mathcal{H}}=-\hat{\vec{\mu}}\cdot\vec{B}=-\hat{\mu}_{z}B=-g\mu_{_{B}}\hat{S}_{z}B
\end{equation}

where g=2, for J=S. The $q$-generalized magnetic moment thermal
average $\langle\hat{\mu}_z\rangle_q$ is:

\begin{equation}\label{momento_medio}
  \langle\hat{\mu}_{z}\rangle_{q}=g\mu_{\textnormal{\tiny{B}}}\langle\hat{S}_{z}\rangle_{q}
  =g\mu_{\textnormal{\tiny{B}}}SB_{\textnormal{\tiny{S}}}^{\textnormal{\tiny{(q)}}}
\end{equation}

where $\langle\hat{S}_z\rangle_q$ is $q$-generalized thermal
average spin operator and
$B_{\textnormal{\tiny{S}}}^{\textnormal{\tiny{(q)}}}$ is the
\emph{generalized Brillouin function}.  This first quantity can be
determined in the Tsallis framework of normalized $q$-expectation
values\cite{PA_261_1998_534}, from:

\begin{equation}\label{spin_medio}
  \langle\hat{S}_{z}\rangle_{q}=\frac{Tr\{\hat{\rho}^{q}\hat{S}_{z}\}}
  {Tr\{\hat{\rho}^{q}\}}
\end{equation}

where $\hat{\rho}$ is the density operator, derived from the
maximization of the entropy. Below, we analyze two different
proposals for the entropy and show  that re-scaling the Lagrange
parameter, the same density operator emerges from both
definitions.

\subsection{Tsallis Entropy}

The entropy of the system is defined
as\cite{JSP_52_1988_479,JPAMG_24_1991_L69,PA_261_1998_534}:

\begin{equation}\label{entropia_tsallis}
\mathcal{S}_{q}=Tr\{\hat{\rho}^{q}\hat{\mathcal{S}}_{q}\}
=\frac{k}{(q-1)}[1-Tr\{\hat{\rho}^{q}\}]
\end{equation}

where,

\begin{equation}\label{operador_entropia}
  \mathcal{\hat{S}}_{q}=-k \ln_{q}\hat{\rho}
\end{equation}

and $k$ a positive constant. The generalized logarithm is defined
as\cite{JPAMG_31_1998_5281}:

\begin{equation}\label{logaritmo_genera}
  \ln_{q}f=(1-q)^{-1}(f^{1-q}-1)
\end{equation}

The corresponding density operator $\hat{\rho}$ can be determined
from the maximization of the $\mathcal{S}_q$ functional, subjected
to $q$-normalized constraint\cite{PA_261_1998_534}:

\begin{equation}\label{energia_interna}
  U_{q}=\frac{Tr\{\hat{\rho}^{q}\hat{\mathcal{H}}\}}
  {Tr\{\hat{\rho}^{q}\}}
\end{equation}

and the normalization of density operator $Tr\{\hat{\rho}\}=1$,
leading to:

\begin{equation}\label{operadordensidade1}
\hat{\rho}=\frac{1}{Z_{q}}\left[1-(1-q)\frac{\beta}{Tr\{\hat{\rho}^{q}\}}
(\hat{\mathcal{H}}-U_{q})\right]^{\frac{1}{1-q}}
\end{equation}

where $Z_q$ is the generalized partition function:

\begin{equation}\label{particaofunction1}
  Z_{q}=Tr\left\{\left[1-(1-q)\frac{\beta}{Tr\{\hat{\rho}^{q}\}}
  (\hat{\mathcal{H}}-U_{q})\right]^{\frac{1}{1-q}}\right\}
\end{equation}

and $\beta$ is the Lagrange parameter associated to internal
energy constraint\cite{PA_261_1998_534,{livro_tsallis}}. In this
case, the entropy has a well defined concavity, for any value of
$q$, being concave for $q>0$, and convex for
$q<0$\cite{JSP_52_1988_479,{JPAMG_24_1991_L69},{PA_261_1998_534},{livro_tsallis}}.

\subsection{Normalized Tsallis Entropy}
If the entropy functional $\mathcal{\hat{S}}_q$ is defined as
proposed by Rajagopal and Abe\cite{PRL_83_1999_1711}:

\begin{equation}\label{entropianaonorma}
  \mathcal{S}_{q}=\frac{Tr\{\hat{\rho}^{q}\hat{\mathcal{S}}_{q}\}}
  {Tr\{\hat{\rho}^{q}\}}=\frac{k}{(q-1)}\left[\frac{1}{Tr\{\hat{\rho}^{q}\}}-1\right]
\end{equation}

the density operator $\hat{\rho}$ can be determined in a similar
way as above, yielding:

\begin{equation}\label{operadordensidade2}
  \hat{\rho}=\frac{1}{Z_{q}}\left[1-(1-q)\beta Tr\{\hat{\rho}^{q}\}
  (\hat{\mathcal{H}}-U_{q})\right]^\frac{1}{1-q}
\end{equation}

\begin{equation}\label{funcaoparticao2}
  Z_{q}=Tr\left\{\left[1-(1-q)\beta Tr\{\hat{\rho}^{q}\}
  (\hat{\mathcal{H}}-U_{q})\right]^\frac{1}{1-q}\right\}
\end{equation}

In this case, the $q$ parameter is restricted to the interval
$0\leq q \leq1$, preserving the entropy
concavity\cite{PRL_83_1999_1711}.

The density operators emerging from both cenarios
(Eq.\ref{operadordensidade1} and \ref{operadordensidade2}) can be
re-written in the form:

\begin{equation}\label{operadordensidade3}
  \hat{\rho}=\frac{1}{Z_{q}'}\left[1-(1-q)\beta^{*}\hat{\mathcal{H}}\right]^\frac{1}{1-q}
\end{equation}

\begin{equation}\label{funcaoparticao3}
  Z_{q}'=Tr\left\{\left[1-(1-q)\beta^{*}\mathcal{\hat{H}}\right]^\frac{1}{1-q}\right\}
\end{equation}

where,
\begin{equation}\label{betaa}
  \beta^{*}=\frac{\beta}{Tr\{\hat{\rho}^{q}\}+(1-q)U_{q}\beta}
\end{equation}

for the case A, and:

\begin{equation}\label{betab}
  \beta^{*}=\frac{\beta}{\frac{1}{Tr\{\hat{\rho}^{q}\}}+(1-q)U_{q}\beta}
\end{equation}

for case B. Here, we suggested that the effective temperature is:

\begin{equation}\label{temp}
  T=\frac{1}{k \beta^{*}}
\end{equation}

and the density operator becomes independent of the initial
entropy functional.

The magnetic behavior of a S-spin system will be analyzed as a
function of the parameter
$x^*=g\mu_{\textnormal{\tiny{B}}}SB\beta^*$. In fact, in terms of
$x^*$, the generalized Brillouin function, is given by:

\begin{equation}\label{brillouin}
  B_{S}^{(q)}=\frac{1}{S}\langle\hat{S}_{z}\rangle_{q}=\frac{1}{S}
  \frac{\sum\limits_{m_{s}=-S}^{+S}m_{s}\left[1+(1-q)x^{*}\frac{m_{s}}{S}\right]^\frac{q}{1-q}}
  {\sum\limits_{m_{s}=-S}^{+S}\left[1+(1-q)x^{*}\frac{m_{s}}{S}\right]^\frac{q}{1-q}}
\end{equation}

It is to be remarked that cut-off
procedure\cite{livro_tsallis,{PLA_256_1999_221},{BJP_29_1999_50},{BJP_29_1999_1}}
implies that those states that do not satisfy the condition:

\begin{equation}\label{cut_off}
  1+(1-q)x^{*}\frac{m_{s}}{S}\geq 0
\end{equation}

must be excluded from the summation. In other words, these states
are assigned with a zero probability amplitude, preserving the
positive definition of the density
operator\cite{PRE_52_1995_R3317}.

For a system constituted by N spins S particles the
$q$-generalized spin operator thermal average
(Eq.\ref{spin_medio}), can be written as:

\begin{equation}\label{spin_operator_multipli}
  \langle\hat{S}_{z}\rangle_{q,N}=
   \frac{\sum\limits_{m_{s}=-NS}^{+NS}Y(m_{s})m_{s}\left[1+(1-q)x^{*}\frac{m_{s}}{S}\right]^\frac{q}{1-q}}
  {\sum\limits_{m_{s}=-NS}^{+NS}Y(m_{s})\left[1+(1-q)x^{*}\frac{m_{s}}{S}\right]^\frac{q}{1-q}}
\end{equation}

where, $Y(m_s)$ is the multiplicity and :

\begin{equation}\label{multiplicidade}
  \sum\limits^{NS}_{m_{s}=-NS}Y(m_{s})=(2S+1)^{N}
\end{equation}

In the particular case of N $1/2$-spins particles, $Y(m_s)$ is
simply given by:

\begin{equation}\label{multiplicidade2}
  Y(m_{s})=\frac{N!}{\left(\frac{N}{2}-m_{s}\right)!\left(\frac{N}{2}+m_{s}\right)!}
\end{equation}

\section{Results and Discussion}

When $q$ is different from unity, general expressions for magnetic
observables are difficult to calculate and interpret, even for
simple systems. This hinders the comparison between theoretical
predictions and experimental results. Numerical methods, on the
other hand, provide a means to directly calculate observables,
allowing a deeper understating of the theoretical results.

Figure 1 displays the generalized Brillouin function
$B_{\textnormal{\tiny{S}}}^{\textnormal{\tiny{(q)}}}$ vs. $x^*$
(Eq.\ref{brillouin}), for different values of $q$ and S=5/2. For
$q$ up to 0.5, a series of kinks appear in the curve. One also
notes that the saturation value
$B^{\textnormal{\tiny{(q)}}}_{\textnormal{\tiny{S}}}|_{\textnormal{\tiny{SAT}}}$
decreases with decreasing $q$. This is more clearly shown in
figure 2(a) and (b), where
$B^{\textnormal{\tiny{(q)}}}_{\textnormal{\tiny{S}}}|_{\textnormal{\tiny{SAT}}}$
is plotted as a function of $q$, for half-integer and integer spin
values, respectively. The behavior of a classical spin is also
included and can be exactly calculated from Eq.\ref{brillouin}
with $S\rightarrow \infty$, giving:

\begin{equation}\label{magsat}
B_{\textnormal{\tiny{S}}}^{\textnormal{\tiny{(q)}}}|_{\textnormal{\tiny{SAT}}}=\frac{1}{2-q}
\end{equation}

The occupation probability (OP), as a function of $x^*$, for each
energy level of S=5/2 are displayed in figure 3(a)(b)(c), for
several values of $q$. Figure 3(d) shown the same quantity for S=2
and q=0.1. We observe that the OP does not vanish for negative
energy levels ($m_s>0$), even for very large values of $x^*$, in
contrast to what happens in the case q=1 (Maxwell-Boltzmann). From
figure 3(c) we can see that, for q=0.1, the OP of the positive
energy states ($m_s<0$) vanish sharply at the same $x^*$ values as
the kinks observed in figure 1. This occurs for q<0.5. Therefore,
we correlate the kinks observed on the magnetization curves to the
lost of occupation of the most energetic states. This is
consequence of Tsallis' cut-off. In fact, from Eq.\ref{cut_off},
the $x^*$ values where the kinks occurs  can be derived,
following:

\begin{equation}\label{xkinks}
  x^{kinks}_{m_{s}}=\frac{S}{|m_{s}|(1-q)}
\end{equation}

It is to be remarked that for a half-integer spin, S+1/2 kinks
occur, and S for an integer spin.

Figure 4(a)(b) display, respectively, the unormalized
(Eq.\ref{entropia_tsallis}) and normalized
(Eq.\ref{entropianaonorma}) entropy $\mathcal{S}_q$, for S=5/2 and
different $q$ values. Note that, for q<0.5, the kinks discussed
before are present. Besides, for any q<1, the entropy does not
vanish in the limit $x^{*^{-1}}\rightarrow 0$. In other words,
even for high field and/or low temperature, the system has a
finite entropy, which prevents a fully magnetized state. These
features are valid for both entropy functionals (section II).

A general expression for the magnetic susceptiblity can be deduced
from Eq.\ref{spin_medio} and Eq.\ref{brillouin}:

\begin{equation}\label{sus}
  \chi_{q}=\lim_{B\to\\0} \left[\frac{\partial\langle\hat{\mu}_z\rangle_q}{\partial B}
  \right]=\frac{C^{(q)}}{T}
\end{equation}

where $C^{(q)}=qC^{(1)}$ is the generalized Curie constant. Figure
5 displays $\chi_q^{-1}$ as a function of $x^{*^{-1}}$ for q=1.0
and 0.8, with S=5/2.

In order to compare the predictions of the proposed model to
experimental data, we must investigate how the magnetic
observables discussed before scales upon increasing the number of
particles in the system. An extensive quantity $\mathcal{F}$
scales as

\begin{equation}\label{limite_termodinamico}
  \mathcal{F}_N=N\mathcal{F}_1
\end{equation}

where N is the system number of particle.

Particularly useful for the purpose of comparison, is the magnetic
susceptibility. Portesi et al.\cite{PRE_52_1995_R3317} and Nobre
et al.\cite{PMB_73_1996_754}, considered the generalized
magnetization on paramagnetic fase of a N $1/2$-spin system in the
unormalized
formalism\cite{JPAMG_24_1991_L69,{PA_261_1998_534},{PRE_52_1995_R3317},
{PMB_73_1996_754},{livro_tsallis}}. They found for $\chi_q$:

\begin{equation}\label{sus_diverge}
  \chi_{q}=\frac{C_{\textnormal{\tiny{S=1/2}}}^{\textnormal{\tiny{(q)}}}}{T}2^{N(1-q)}
\end{equation}

where,

\begin{equation}\label{curie_generalizada}
  C^{\textnormal{\tiny{(q)}}}_{\textnormal{\tiny{S=1/2}}}=\frac{(g \mu_{\textnormal{\tiny{B}}})^2}{4k}Nq
\end{equation}

and $T=1/k\beta$ is the temperature of the system. Here, $\beta$
is the usual Lagrange parameter associated to the non-normalized
constraint of internal energy.

Therefore, as N increases, the magnetic susceptibility becomes
infinity for q<1 and zero for q>1. In other words, there is no
linearity among $\chi_q$ and N. This phenomenon is called
\emph{dark magnetism}, in analogy to the cosmological concept of
\emph{dark matter} [\cite{PRE_52_1995_R3317} and references there
in].

S. Martínez et al.\cite{PA_282_2000_193}, analyzed the
paramagnetic behavior of the same N $1/2$-spin system, however,
within the normalized
formalism\cite{PA_261_1998_534,{livro_tsallis}}, and found a
similar result as Portesi and Nobre for $\chi_q$.

Our proposal is that the paramagnetic behavior of a non-extensive
N $1/2$-spin system should be analyzed in the normalized
formalism, using the density operator described in
Eq.\ref{operadordensidade3}, taking $\beta^*$, instead $\beta$,
inversely proportional to the system temperature T. By doing so,
the generalized paramagnetic susceptibility becomes:

\begin{equation}\label{sus_nossa}
  \chi_{q}=\frac{C^{\textnormal{\tiny{(q)}}}_{\textnormal{\tiny{S=1/2}}}}{T}
\end{equation}

which does not diverge as N increases, and is indeed proportional
to N. Besides, this result is independent of the choice for the
functional entropy.

\section{Possible Connection to Experimental Results}

Amaral et al.\cite{JMMM_226_2001_837} discussed the magnetic
behavior of manganese oxides, namely,
La$_{0.67}$Ca$_{0.33}$MnO$_{3}$ and verified steps on the curve of
M$^{-1}$ vs. T in the paramagnetic phase, as shown in figure 6(a).
These steps are analogue to those shown on figure 6(b), that
represents also the inverse of magnetization as a function of
$x^{*{-1}}$, for q=0.1, and encourage the idea of manganites as
non-extensive objects\cite{EL_58_2002_42}. The authors relate the
change in the slope of the curve as an indication of cluster
formation, which change the effective moment of Mn ions. These
clusters could give rise to fractal structures, as discussed by
Dagotto\cite{PR_344_2001_1}, and therefore are in accordance to
the ideas discussed here.

In addition, Hébert et al.\cite{JSSC_165_2002_6} found
magnetization curves in
Pr$_{0.5}$Ca$_{0.5}$Mn$_{0.95}$Ga$_{0.05}$O$_{3}$ that are also in
qualitative agreement to those curves in figure 1.

\section{Conclusion}
In summary, we investigated the properties of a paramagnetic
S-spin system under Tsallis generalized statistics, on the
normalized formalism. For q<0.5, a series of kinks appears in
magnetization and entropy. This effect is a direct consequence of
a peculiar occupation probability, as a result of Tsallis'
cut-off, where the positive energy states ($m_s<0$) vanish
sharply. Additionally, the negative energy states ($m_s>0$) share
non-zero occupation probability, preventing a fully magnetized
state and the saturation magnetization decreases with decreasing
$q$. We present evidences based on experimental results of Amaral
et al.\cite{JMMM_226_2001_837} and Hébert et
al.\cite{JSSC_165_2002_6} which add and support our previous
publication\cite{EL_58_2002_42},  where manganites are suggested
to be magnetically non-extensive objects.

\begin{acknowledgments}
One of us, M.S.R., acknowledges CAPES, Brazil, for financial
support on Aveiro University. I.S.O. acknowledges FAPERJ, the Rio
de Janeiro Funding Agency.
\end{acknowledgments}

\section{Captions}
Figure 1: Generalized Brillouin function for several $q$ values
and S=5/2, as a function of $x^*$.

Figure 2: Saturation magnetization
$B^{\textnormal{\tiny{(q)}}}_{\textnormal{\tiny{S}}}|_{\textnormal{\tiny{SAT}}}$
as a $q$ function, for (a)half-integer and (b)integer S values.

Figure 3: Probability of occupation of each energy level, for
(a)(b)(c) S=5/2 and (d) S=2, as a function of $x^*$, and various
values of $q$.

Figure4: (a) Unormalized (Eq. \ref{entropia_tsallis}) and (b)
normalized (Eq. \ref{entropianaonorma}) entropy $\mathcal{S}_q$,
for S=5/2 and different $q$ values, as a function of $x^{*^{-1}}$.

Figure5: Inverse of susceptibility for (a) q=1.0 and 0.8, for
S=5/2, as a $x^{*^{-1}}$ function.

Figure6: (a) Temperature dependence of the inverse susceptibility
for the manganite La2/3Ca1/3MnO3, above the Curie temperature
(~267 K), at low magnetic field. (b)Inverse of magnetization
(Eq.\ref{brillouin}) as a function of $x^{*{-1}}$, for q=0.1

\end{document}